\documentclass[pdflatex,sn-mathphys-num]{sn-jnl}

\usepackage[utf8]{inputenc}
\usepackage{amssymb}
\usepackage{physics}

\begin{document}
\title[Quantum defects of Rydberg excitons]{Quantum defects of Rydberg
  excitons in cuprous oxide: A semiclassical spherical model}
\author[1]{\fnm{Jan} \sur{Ertl}}\nomail
\author[1]{\fnm{Patric} \sur{Rommel}}\nomail
\author*[1]{\fnm{Jörg} \sur{Main}}\email{main@itp1.uni-stuttgart.de}
\affil[1]{\orgdiv{Institut für Theoretische Physik I},
  \orgname{Universität Stuttgart}, \orgaddress{70550 Stuttgart}, \country{Germany}}

\abstract{
Excitons, i.e.\ the bound states of an electron and a positively
charged hole are the solid state analogue of the hydrogen atom.
As such they exhibit a Rydberg series, which in cuprous oxide has been
observed up to high principal quantum numbers by T.~Kazimierczuk et al.\
[Nature 514, 343 (2014)].  In this energy regime the quantum
mechanical properties of the system can be understood in terms of
classical orbits by the application of semiclassical techniques.
In fact the first theoretical explanation of the spectrum of the
hydrogen atom within Bohr's atomic model was a semiclassical one using
classical orbits and a quantization condition for the angular momentum.
Contrary to the hydrogen atom, the degeneracy of states with the same
principal quantum number $n$ is lifted in exciton spectra.  This is
similar to the situation in alkali atoms, where these splittings are
caused by the interaction of the excited electron with the ionic core.
For excitons in cuprous oxide, these splittings occur due to the
influence of the complex band structure of the crystal.  Using an
adiabatic approach and analytically derived energy surfaces, we
develop a semiclassical spherical model and determine, via
semiclassical torus quantization, the quantum defects of various
angular momentum states.
}

\maketitle

\section{Introduction}
\label{sec:introduction}
At the end of the 19th century it became evident that classical
physics must be extended to describe phenomena observed in microscopic
systems, such as the distinct lines of atomic spectra, where the
energy is quantized.
The first quantitative description was introduced by Balmer in 1885
providing a formula correctly reproducing the lines in the range of
visible light for the hydrogen atom~\cite{balmer1885notiz}.
This formula was extended by Rydberg~\cite{rydberg1890on}
predicting the existence of other series in addition to the Balmer
series which only includes transitions from or to states with
principal quantum number $n=2$.
The Rydberg formula was confirmed after the Lyman
series~\cite{lyman1906spectrum} ($n=1$) and the Paschen
series~\cite{paschen1908kenntnis} ($n=3$) were found.
The first theoretical explanation for the hydrogenic spectra was
given by Bohr~\cite{bohr1913constitution} by postulating that the
electron can only move on certain classical orbits fulfilling a
quantization condition for the angular momentum, which allows for a
derivation of the empirically known Rydberg formula.
Sommerfeld extended the quantization of only circular orbits to
ellipses via a quantization condition and quantum numbers for all
three spherical coordinates~\cite{sommerfeld1916quantentheorie}.

The work by Bohr and Sommerfeld preceded the modern formulation 
of quantum mechanics where the particles are described by a wave 
function. A semiclassical treatment of the wave functions reveals
that one also has to account for the caustics when performing 
a semiclassical torus quantization.
To this aim one has to consider the Maslov index counting the number
of reflections and turning points along independent paths on the torus.
The resulting Einstein-Brillouin-Keller (EBK) quantization
condition~\cite{einstein1917quantensatz,Brillouin1926remarques,keller1958corrected,maslov2001semi}
for the corresponding action variables provides good results for a
wide range of integrable and even near-integrable systems.
However, as already noticed by Einstein~\cite{einstein1917quantensatz}
the EBK quantization cannot be applied to non-integrable or chaotic
systems.
Semiclassical theories for such systems have been derived in the
1970's by Gutzwiller~\cite{gutzwiller1967phase,gutzwiller1971periodic,Gutzwiller1990Chaos,brack1997semiclassical}.

The hydrogen-like model is also appropriate to deal with
non-hydrogenic atoms with one highly excited electron.
For these Rydberg atoms the distance of the electron to the nucleus
and the other electrons is large, meaning that the ionic core
can approximately be treated as one entity with net charge $+e$.
For the hydrogen atom all states with a given principal 
quantum number $n$ are degenerate.
For non-hydrogenic atoms this degeneracy is partly lifted.
In the classical picture this can be understood when looking at the
corresponding orbits.
For states with low angular momentum quantum numbers $l$ the orbits of
the excited electron come closer to the core, which in this region is
no longer fully screened by the other electrons.
Thus, the excited electron experiences a stronger Coulomb interaction,
lowering the binding energy.
To capture these deviations from the hydrogen spectrum, Rydberg himself
introduced a correction scheme to modify the expressions for the wave
number by subtracting a non-integer quantum defect $\delta_{n,l}$ from
the principal quantum number and substituting $n$ by
$n^*=n-\delta_{n,l}$ in the Rydberg formula~\cite{rydberg1890on}.
As the screening of the core becomes stronger for increasing angular
momentum quantum number, the quantum defects typically decrease with
increasing $l$.

Excitons consisting of a positively charged hole and a negatively charged
electron are the solid-state analogue of the hydrogen atom.
For most systems only the lowest principal quantum numbers
are experimentally accessible~\cite{Baldini1962Ultraviolet,Sturge1962Optical,mueller2018exciton}.
This was also true for first experiments on cuprous oxide~\cite{Gross1956Optical},
however, this has changed since the observation of high Rydberg
excitons with principal quantum numbers up to $n=25$ in the seminal
work by Kazimierczuk et al.~\cite{Kazimierczuk2014Giant}.
Nowadays, a Rydberg series of $p$ states can be traced up to 
$n=30$~\cite{Versteegh2021Giant}. 
The rough features of the spectra can be explained within a simple
hydrogen-like model by the Coulomb interaction between electron and
hole, however, a detailed description requires the consideration of
band-structure effects of the crystal~\cite{luttinger1955motion,luttinger1956quantum,Schweiner2016Impact},
which cause the reduction of the spherical symmetry to the cubic point
group $O_{\mathrm{h}}$~\cite{koster1963properties}.
Additional central-cell corrections become important when electron and
hole come close to each other~\cite{Schweiner2017Even}.
In the 1950's, Hermann Haken provided fundamental contributions to
this topic, including the Haken-potential~\cite{Haken1958Theorie,Haken1956Quantentheorie,Haken1956Kopplung,Haken1957,Haken1958Behandlung,KuperWhitfield1963}.
Later, he founded the field of synergetics dealing with complex systems.
Here, microscopic degrees of freedom are adiabatically eliminated, 
allowing for a description of the system in terms of a small number of
order parameters.
We have tried to bridge the two topics by utilizing an adiabatic
approach to the spin degrees of freedom in the exciton
dynamics~\cite{Ertl2020Classical}.
This leads to a classical description of the nonlinear exciton
dynamics via energy surfaces in momentum space and the observation of
a transition from a regular to chaotic motion ~\cite{ertl2024rydberg}.
The classical periodic orbits have been linked to the quantum spectra
of the system by application of semiclassical
techniques~\cite{Ertl2022Signatures,ertl2024classical}.
A review on experimental and theoretical results on the yellow
exciton series in cuprous oxide is given in a recent article by
Heckötter et al.~\cite{Heckoetter2025energy}.
The highly excited $p$ states of the yellow exciton series
can be described by a Rydberg formula with quantum
defects~\cite{Kazimierczuk2014Giant}.
This concept can also be applied to other angular momentum
states~\cite{Schoene2016Coupled,Schoene2016Deviations}.
Uihlein et al.~\cite{Uihlein1981Investigation} adjusted a spherical
model to experimental spectra.

Our goal in this article is to derive quantum defects of the yellow
exciton series from a semiclassical spherical model.
We start, in Sec.~\ref{sec:Hamiltonian}, from the full model including
the cubic contributions and apply, in Sec.~\ref{sec:adiabat_appr},
an adiabatic approach to obtain energy surfaces for a
classical description of the exciton dynamics.
Analytical formulas for these energy surfaces are derived in
Sec.~\ref{sec:analytic_expr}.
In Sec.~\ref{sec:spherical_model} we obtain spherically symmetric
energy surfaces using numerical and analytical methods, which are the
basis for the semiclassical EBK quantization in Sec.~\ref{sec:EBK}.
Considering the probability distribution of the exciton in momentum
space leads to order parameters describing the different angular
momentum manifolds.
A conclusion and outlook are given in Sec.~\ref{sec:conclusion}.

\section{Theory}
\label{sec:theory}
Excitons in a semiconductor are formed when an electron is excited
from the valence band into the conduction band.
Instead of taking into account the interaction of the excited electron
with the solid and all its constituents, one treats the missing hole in
the valence band as positively charged quasi particle, which can, via
the Coulomb interaction with the electron, form a hydrogen-like bound
state, the exciton.

\subsection{Hamiltonian for excitons in cuprous oxide}
\label{sec:Hamiltonian}
The Hamiltonian for a theoretical description of excitons is given as
\begin{equation}
  H=H_{\mathrm{e}}\left(\boldsymbol{p}_{\mathrm{e}}\right)+H_{\mathrm{h}}\left(\boldsymbol{p}_{\mathrm{\mathrm{h}}}\right) - \frac{e^{2}}{4\pi\varepsilon_{0}\varepsilon}\frac{1}{\left|\boldsymbol{r}_{\mathrm{e}}-\boldsymbol{r}_{\mathrm{h}}\right|}\, ,
  \label{eq:Hpeph}
\end{equation}
with the kinetic energy of the electron
$H_{\mathrm{e}}(\boldsymbol{p}_{\mathrm{e}})$ and the
kinetic energy of the hole
$H_{\mathrm{h}}(\boldsymbol{p}_{\mathrm{h}})$.
The Coulomb interaction is screened by the permittivity $\varepsilon$
due to the crystal environment.
In cuprous oxide the kinetic energy for the hole follows a parabolic
dispersion~\cite{Schoene2016Deviations,Schoene2016Coupled,Schweiner2016Impact}
\begin{equation}
  H_{\mathrm{e}}(\boldsymbol{p}_{\mathrm{e}})
  = E_{\mathrm{g}}+\frac{\boldsymbol{p}_{\mathrm{e}}^{2}}{2m_{\mathrm{e}}} \, ,
\end{equation}
with the gap energy $E_{\mathrm{g}}$ and the effective electron mass
$m_{\mathrm{e}}$.
The kinetic energy of the hole is more complicated, because the
uppermost valence band has a threefold degeneracy at the
$\Gamma$ point.
This situation can be modeled by introducing a quasispin $I=1$.
Due to the spin-orbit coupling of the hole spin $S_{\mathrm{h}}=1/2$
and the quasispin
\begin{equation}
  H_{\mathrm{so}}=\frac{2}{3}\Delta
  \left(1+\frac{1}{\hbar^2}\boldsymbol{I}\cdot\boldsymbol{S}_{\mathrm{h}}\right)\, ,
\label{eq:H_SO}
\end{equation}
the valence band splits, leading to a yellow series and a green series
with heavy holes (hh) and light holes (lh).
The Hamiltonian for the hole in the valence band needs to
consider all terms including momenta, quasispin and hole spin
in line with the cubic $O_{\mathrm{h}}$ symmetry of the crystal.
Up to second order of the momenta the Hamiltonian for the hole
reads~\cite{Lipari1977Theory,luttinger1956quantum,Schoene2016Coupled,Schoene2016Deviations,Schweiner2016Impact}
\begin{align}
  H_{\mathrm{h}}(\boldsymbol{p}) & = H_{\mathrm{so}}+\left(1/2\hbar^{2}m_{0}\right)\left\{ \hbar^{2}\left(\gamma_{1}+4\gamma_{2}\right)\boldsymbol{p}^{2}\right.
     +2\left(\eta_{1}+2\eta_{2}\right)\boldsymbol{p}^{2}\left(\boldsymbol{I}\cdot\boldsymbol{S}_{\mathrm{h}}\right)\nonumber \\
   &  -6\gamma_{2}\left(p_{1}^{2}\boldsymbol{I}_{1}^{2}+\mathrm{c.p.}\right)-12\eta_{2}\left(p_{1}^{2}\boldsymbol{I}_{1}\boldsymbol{S}_{\mathrm{h}1}+\mathrm{c.p.}\right)\nonumber \\
   &  -12\gamma_{3}\left(\left\{ p_{1},p_{2}\right\} \left\{ \boldsymbol{I}_{1},\boldsymbol{I}_{2}\right\} +\mathrm{c.p.}\right)\nonumber \\
   &  \left.-12\eta_{3}\left(\left\{ p_{1},p_{2}\right\} \left(\boldsymbol{I}_{1}\boldsymbol{S}_{\mathrm{h}2}+\boldsymbol{I}_{2}\boldsymbol{S}_{\mathrm{h}1}\right)+\mathrm{c.p.}\right)\right\}\, , 
\label{eq:Hh}
\end{align}
with the Luttinger parameters $\gamma_i, \eta_i$. 
While in principle higher orders of the momenta are possible, their
contributions can be neglected for our purposes~\cite{Schweiner2017Even}.
In our classical and semiclassical approach we also ignore
central-cell corrections, which affect the even-parity
excitons~\cite{Schweiner2017Even}.
Using relative and center-of-mass coordinates and neglecting the
center-of-mass motion, the full Hamiltonian for excitons in cuprous
oxide reads
\begin{align}
  H & = E_{\mathrm{g}}+ H_{\mathrm{so}}+\left(1/2\hbar^{2}m_{0}\right)\left\{ \hbar^{2}\left(\gamma'_{1}+4\gamma_{2}\right)\boldsymbol{p}^{2}\right.
     +2\left(\eta_{1}+2\eta_{2}\right)\boldsymbol{p}^{2}\left(\boldsymbol{I}\cdot\boldsymbol{S}_{\mathrm{h}}\right)\nonumber \\[1ex]
   &  -6\gamma_{2}\left(p_{1}^{2}\boldsymbol{I}_{1}^{2}+\mathrm{c.p.}\right)-12\eta_{2}\left(p_{1}^{2}\boldsymbol{I}_{1}\boldsymbol{S}_{\mathrm{h}1}+\mathrm{c.p.}\right)
     -12\gamma_{3}\left(\left\{ p_{1},p_{2}\right\} \left\{ \boldsymbol{I}_{1},\boldsymbol{I}_{2}\right\} +\mathrm{c.p.}\right)\nonumber \\
   &  \left.-12\eta_{3}\left(\left\{ p_{1},p_{2}\right\} \left(\boldsymbol{I}_{1}\boldsymbol{S}_{\mathrm{h}2}+\boldsymbol{I}_{2}\boldsymbol{S}_{\mathrm{h}1}\right)+\mathrm{c.p.}\right)\right\}
     - \frac{e^{2}}{4\pi\varepsilon_{0}\varepsilon}\frac{1}{\left|\boldsymbol{r}_{\mathrm{e}}-\boldsymbol{r}_{\mathrm{h}}\right|}\, ,
\label{eq:H}
\end{align}
with $\gamma'_{1}=\gamma_{1}+m_{0}/m_{\mathrm{e}}$.  
The Hamiltonian~\eqref{eq:H} can be split into terms with spherical
symmetry and terms with reduced cubic symmetry by using irreducible
tensors~\cite{Baldereschi1973Spherical,edmonds1996angular,Broeckx1991Acceptor,Schweiner2016Impact},
\begin{align}
H & = E_{\mathrm{g}}-\frac{e^{2}}{4\pi\varepsilon_{0}\varepsilon}\frac{1}{r}+\frac{2}{3}\Delta\left(1+\frac{1}{\hbar^{2}}I^{(1)}\cdot S_{\mathrm{h}}^{(1)}\right)
   +\frac{\gamma'_{1}}{2\hbar^{2}m_{0}}\Bigg[\hbar^{2}p^{2}-\frac{\mu'}{3}P^{(2)}\cdot I^{(2)}\nonumber \\
  &+\frac{\delta'}{3}\left(\sum_{k=\pm4}\left[P^{(2)}\times
   I^{(2)}\right]_{k}^{(4)}
   +\frac{\sqrt{70}}{5}\left[P^{(2)}\times I^{(2)}\right]_{0}^{(4)}\right)\Bigg]
   +\frac{3\eta_{1}}{\hbar^{2}m_{0}}\Bigg[\frac{1}{3}p^{2}\left(I^{(1)}\cdot
    S_{\mathrm{h}}^{(1)}\right) \nonumber \\
    &-\frac{\nu}{3}\, P^{(2)}\cdot
   D^{(2)}
   +\frac{\tau}{3}\,\left(\sum_{k=\pm4}\left[P^{(2)}\times
      D^{(2)}\right]_{k}^{(4)}+\frac{\sqrt{70}}{5}\left[P^{(2)}\times
      D^{(2)}\right]_{0}^{(4)}\right)\Bigg] \, .
\label{eq:H0}
\end{align}
For the tensor operators we use the same convention as
in Refs.~\cite{Broeckx1991Acceptor,Schweiner2016Impact}.
The coefficients
\begin{equation}
  \mu'=\frac{6\gamma_{3}+4\gamma_{2}}{5\gamma'_{1}}=0.0586 \, , \quad
  \nu=\frac{6\eta_{3}+4\eta_{2}}{5\eta_{1}}=2.167
\label{eq:mu}
\end{equation}
give the strength of the spherical contributions, whereas the strength
of the cubic contributions is characterized by
\begin{equation}
  \delta'=\frac{\gamma_{3}-\gamma_{2}}{\gamma'_{1}}=-0.404 \, , \quad
  \tau=\frac{\eta_{3}-\eta_{2}}{\eta_{1}}=1.500\, .
\label{eq:delta}
\end{equation}
For the calculation of quantum defects in Sec.~\ref{sec:results} we
use a spherical model, however, contributions of the cubic terms in
Eq.~\eqref{eq:H0} will be considered in a semiclassical picture by
averaging the energy surfaces, derived in Sec.~\ref{sec:analytic_expr},
over the spherical angle coordinates.

\subsection{Adiabatic approach to the exciton dynamics}
\label{sec:adiabat_appr}
In this article we use a semiclassical picture to compute the quantum
defects of Rydberg excitons in cuprous oxide.
For the classical exciton dynamics we resort to the adiabatic approach
presented in Ref.~\cite{Ertl2020Classical}.
The basic idea is that for Rydberg excitons with principal quantum
numbers $n\gtrsim 3$ the energy spacing between adjacent Rydberg
levels is small compared to the spacings caused by the spin-orbit
coupling, and thus the exciton dynamics in the coordinate and momentum
space is much slower than the spin dynamics.
In analogy to the Born-Oppenheimer approximation for molecules, this
allows for a classical or semiclassical treatment of the exciton
dynamics, where the coordinate and momentum operators in the
Hamiltonian~\eqref{eq:H} are replaced by classical variables and only
the spin degrees of freedom are considered quantum mechanically.
The energy surfaces are obtained by diagonalizing the
Hamiltonian $H_{\mathrm{band}}$ defined by the Hamiltonian in
Eq.~\eqref{eq:H0} excluding the gap energy $E_{\mathrm{g}}$, the term
$\gamma'_{1}\boldsymbol{p}^2/(2m_0)$, and the Coulomb potential,
for given values of the classical momenta
$\boldsymbol{p}$ in the basis $\ket{m_I, m_{S_{\rm h}}}$
for the quasispin and hole spin degrees of freedom, with $m_I=0,\pm 1$
and $m_{S_{\mathrm h}}=\pm 1/2$.
This yields a $(6 \times 6)$-dimensional eigenvalue problem
\begin{equation}
  \boldsymbol{H}_{\mathrm{band}}(\boldsymbol{p})\boldsymbol{c}
  = W_k(\boldsymbol{p})\boldsymbol{c}
\label{eq:H6x6}
\end{equation}
for energy surfaces $W_k(\boldsymbol{p})$.
In general, the eigenvalues of a $6 \times 6$ matrix cannot be found
analytically, however, solutions can be obtained numerically, e.g., by
using an appropriate \textit{LAPACK} routine \cite{anderson1999lapack}.
It should be noted that the eigenvalues of the eigenvalue
problem~\eqref{eq:H6x6} are twofold degenerate, i.e., there are only
three independent energy surfaces, where the lowest one belongs to the
yellow exciton series and the two upper surfaces are related to the
green series with heavy and light hole.
This is due to the Kramers degeneracy in time-reversal symmetric
systems with a half-integer total spin~\cite{kramers1930general}.
The calculation of the energy surfaces by numerical diagonalizations of
the $6\times 6$ matrices is rather time-consuming.
Furthermore, for the computation of classical exciton orbits by
solving Hamilton's equations of motion and their stability analysis
first and second derivatives of the energy surfaces
$W_k(\boldsymbol{p})$ are required.
In addition, the energy surfaces are equivalent to the electronic 
band structure.
Therefore, analytical expressions for the energy surfaces are highly
desirable~\cite{schone2017optical}.
Indeed, as will be shown below, it is possible to exploit Kramers'
degeneracy to find analytical expressions for the energy surfaces.

\subsection{Analytical expressions for the energy surfaces}
\label{sec:analytic_expr}
The starting point for the derivation of analytical expressions for
the energy surfaces is the characteristic polynomial of the $6\times 6$
matrix in Eq.~\eqref{eq:H6x6}, which takes the form
\begin{equation}
  \chi(\lambda)=  \lambda^6+c_5 \lambda^5+c_4 \lambda^4+c_3 \lambda^3+c_2 \lambda^2+c_1 \lambda+c_0 \, .
\label{eq:Characteristic_Polynomial_c}
\end{equation}
In general, there are no analytical formulas for the roots of a
polynomial of degree five or higher.
However, for the special case of excitons in cuprous oxide analytical
expressions can be obtained by exploiting Kramers' degeneracy.
Since every energy surface is twofold degenerate, the characteristic
polynomial~\eqref{eq:Characteristic_Polynomial_c} can be expressed as
\begin{equation}
    \left[W_\mathrm{y}(\boldsymbol{p})-\lambda\right]^2\left[W_\mathrm{g,hh}(\boldsymbol{p})-\lambda\right]^2\left[W_\mathrm{g,lh}(\boldsymbol{p})-\lambda\right]^2=0 \, .
\label{eq:Characteristic_Polynomial_W}
\end{equation}
By comparing the left hand side of Eq.~\eqref{eq:Characteristic_Polynomial_W}
with Eq.~\eqref{eq:Characteristic_Polynomial_c} and using the abbreviations
$W_3=W_{\mathrm{y}}(\boldsymbol{p})$, $W_2=W_{\mathrm{g,hh}}(\boldsymbol{p})$, and
$W_1=W_{\mathrm{g,lh}}(\boldsymbol{p})$ the following system of
equations for the energy surfaces is obtained:
\begin{subequations}
\begin{align}
c_5&=-2 \left( W_1+W_2+W_3 \right) \, , \label{eq:c5}\\
c_4&= \left( W_1+W_2+W_3 \right)^2
 +2 W_1 W_2+2 W_1 W_3+2 W_2 W_3 \, ,\label{eq:c4} \\
c_3&= -2 ( W_1^2 W_2 + W_1^2 W_3 + W_1 W_2^2 \nonumber \\
&+ W_1 W_3^2 + W_2^2 W_3 + W_2 W_3^2 + 4 W_1 W_2 W_3 ) \, , \label{eq:c3} \\
c_2&=W_1^2 W_2^2 + W_1^2 W_3^2 + W_2^2 W_3^2
 +4 \left( W_1^2 W_2 W_3+ W_1 W_2^2 W_3 + W_1 W_2 W_3^2 \right) \, , \\
c_1&=-2 \left(W_1^2 W_2^2 W_3 + W_1^2 W_2 W_3^2 + W_1 W_2^2 W_3^2 \right) \, , \\
c_0&=W_1^2 W_2^2 W_3^2 \, .
\end{align}
\end{subequations}
Since there are six equations for three potential surfaces, this
system is overdetermined.
However, these equations are not independent.
To obtain a solution for the energy surfaces it is therefore
sufficient to consider the first three equations
\eqref{eq:c5}--\eqref{eq:c3}.
Using these expressions one arrives at a polynomial of degree three
for the energy surfaces that needs to vanish,
\begin{equation}
W_k^3 + \frac{c_5}{2} W_k^2+ \frac{4 c_4-c_5^2}{8} W_k+ \frac{8 c_3-4 c_4 c_5+c_5^3}{16}=0 \, .
\label{eq:Polynomial_3}
\end{equation}
The solutions of Eq.~\eqref{eq:Polynomial_3} can be found by
applying Cardano's formula, which for real-valued solutions takes 
the form~\cite{bewersdorff2019algebra}
\begin{align}
  W_k= -\frac{c_5}{6}
  +  \sqrt{-\frac{4 p_c}{3}} \cos \left[ \frac{1}{3} \arccos \left(-\frac{q_c}{2} \sqrt{-\frac{27}{p_c^3}}\right) -\frac{2 \pi \left(k-1\right)}{3} \right]\, ,
\label{eq:Wk}
\end{align}
where $p_c$ and $q_c$ are given by the coefficients of the characteristic
polynomial $c_i$, i.e.
\begin{align}
  p_c &= \frac{c_4}{2}-\frac{5 c_5^2}{24} \; ,\quad
  q_c = \frac{c_3}{2}-\frac{c_4 c_5}{3}+\frac{5 c_5^3}{54} \, .
\end{align}
To evaluate the energy surfaces in Eq.~\eqref{eq:Wk} we need to insert
the expressions for $c_5$, $p_c$, and $q_c$ in terms of the momenta
$\boldsymbol{p}$ and the material parameters of cuprous oxide.
The following expressions are obtained:
\begin{subequations}
\label{eq:cpq_coefs}
\begin{align}
  c_5 &= -4 \Delta \, ,\\
  p_c &= a_0 +
a_1 (p_1^2+p_2^2+p_3^2) +
a_2 (p_1^2+p_2^2+p_3^2)^2+ 
a_3 (p_1^2 p_2^2+p_1^2 p_3^2+p_2^2 p_3^2)
 \, ,\\
  q_c &= b_0+
b_1 (p_1^2+p_2^2+p_3^2)+
b_2 (p_1^2+p_2^2+p_3^2)^2+
b_3 (p_1^2+p_2^2+p_3^2)^3\nonumber \\ &+
b_4 (p_1^2 p_2^2+p_1^2 p_3^2+p_2^2 p_3^2)+
b_5 (p_1^2+p_2^2+p_3^2) (p_1^2 p_2^2+p_1^2 p_3^2+p_2^2 p_3^2)\nonumber\\ &+
b_6 (p_1^2 p_2^2 p_3^2) \, ,
\end{align}
\end{subequations}
with the constant coefficients
\begin{subequations}
\label{eq:a_coefs}
\begin{align}
a_0 	&=   - \frac{\Delta^2}{3} \, ,\\
a_1	&=- \frac{\Delta \eta_1}{m_0}  \, , \\
a_2	&=-\frac{3}{100 m_0^2} \left\{\gamma_1'^2 \left(6 \delta' - 5 \mu'\right)^2 + 
  \eta_1^2 \left[25 + 2 \left(5 \nu - 6 \tau\right)^2\right]\right\} \, ,\\
a_3	&=\frac{9}{5m_0^2}	\left[ \gamma_1'^2 (\delta'^2 - 5 \delta' \mu') + 2 \eta_1^2 \tau (-5 \nu + \tau)\right] \, ,
\end{align}
\end{subequations}
and
\begin{subequations}
\label{eq:b_coefs}
\begin{align}
b_0	&= \frac{2 \Delta^3}{27} 
 \, ,\\
b_1	&= \frac{\Delta^2 \eta_1}{3 m_0} \, ,\\
b_2	&=\frac{\Delta \eta_1}{50 m_0^2} \left[ 2 \gamma_1' (6 \delta' - 5 \mu') (5 \nu - 6 \tau) 
  - \eta_1 (-5 + 5 \nu - 6 \tau) (5 + 5 \nu - 6 \tau)\right]
 \, ,\\
b_3	&= \frac{1}{500 m_0^3} \big[\gamma_1'^3 (6 \delta' - 5 \mu')^3 - 
  3 \eta_1^2 \gamma_1' (6 \delta' - 5 \mu') (-10 + 5 \nu - 6 \tau) (5 \nu - 6 \tau) \nonumber\\ &- 
  \eta_1^3 (-5 + 10 \nu - 12 \tau) (5 + 5 \nu - 6 \tau)^2\big]
 \, ,\\
b_4	&= -\frac{6 \Delta \eta_1}{5 m_0^2}
\big[\delta' \gamma_1' (5 \nu - 2 \tau) 
  + \tau (5 \gamma_1' \mu' + 5 \eta_1 \nu - \eta_1 \tau)\big]
 \, ,\\
b_5	&=\frac{9}{50 m_0^3}   \Big\{ -6 \delta'^3 \gamma_1'^3 + 35 \delta'^2 \gamma_1'^3 \mu'  \nonumber \\
&  - \eta_1^2 \tau \left[2 \eta_1 (5 + 5 \nu - 6 \tau) (5 \nu - \tau) + 
     5 \gamma_1' \mu' (10 - 10 \nu + 7 \tau)\right] \nonumber \\
     &+ \delta' \gamma_1' \left[-25 \gamma_1'^2 \mu'^2 + 
     \eta_1^2 (25 (-2 + \nu) \nu + 10 (2 - 7 \nu) \tau + 18 \tau^2)\right] \Big\}
 \, ,\\
b_6	&= -\frac{27}{10 m_0^3} (\delta' \gamma_1' - \eta_1 \tau) \big[2 \delta'^2 \gamma_1'^2 + 
  \eta_1 \tau (15 \gamma_1' \mu' - 30 \eta_1 \nu - 4 \eta_1 \tau) \nonumber\\ &+ 
   \delta' \gamma_1' (15 \gamma_1' \mu' + 2 \eta_1 \tau)\big]
 \, .
\end{align}
\end{subequations}
We finally obtain the Hamiltonian
\begin{equation}
  H = E_{\mathrm{g}}+\frac{\gamma'_{1}}{2 m_0} \boldsymbol{p}^2
  - \frac{e^{2}}{4\pi\varepsilon_{0}\varepsilon}\frac{1}{r} + W_k(\boldsymbol{p})
\end{equation}
for a classical description of the exciton dynamics via three distinct
analytically given energy surfaces $W_k(\boldsymbol{p})$ in momentum
space.
To investigate the yellow exciton series we can now choose the lowest
energy surface $W_{\mathrm{y}}{\left(\boldsymbol{p}\right)}$ and integrate
Hamilton's equations of motion
\begin{equation}
  \dot r_i = \frac{\gamma'_1}{m_0} p_i + \frac{\partial
              W_{\mathrm{y}}(\boldsymbol{p})}{\partial p_i} \; , \quad
  \dot p_i = -\frac{e^2}{4\pi\varepsilon_0\varepsilon}
              \frac{r_i}{|\boldsymbol{r}|^3} \, . 
\label{eq:eom}
\end{equation}
For our calculations we use exciton-Hartree units obtained by setting
$\hbar = e = m_{\mathrm{0}} / \gamma_1' = 1 / (4\pi\varepsilon_{\mathrm{0}} \varepsilon) = 1$,
and the material parameters of cuprous oxide as given in
Eqs.~\eqref{eq:mu} and \eqref{eq:delta}.

\section{Results and discussion}
\label{sec:results}
The energy surfaces $W_k(\boldsymbol{p})$ in Eq.~\eqref{eq:Wk}
incorporate the complex band structure of the system and thus model
the complex interaction of the exciton with the other electrons of the
solid.
Considering the reduced $O_{\mathrm{h}}$ symmetry of the system the
exciton dynamics described by the equations of motion~\eqref{eq:eom}
exhibits a mostly regular dynamics on tori for the yellow exciton
series and a strong transition to chaotic dynamics for the green
exciton series~\cite{ertl2024classical,ertl2024rydberg}.
Here, our goal is to obtain the quantum defects of the yellow excitons
in cuprous oxide.
While a hydrogen-like model due to the high $O(4)$ symmetry of the
Coulomb problem is not capable of producing a splitting of the
different $l$ states, this can be achieved using a model with
spherical symmetry.
To this aim, in a first first step, we derive spherically symmetric
energy surfaces, which are then used, in a second step, for
semiclassical EBK quantization.

\subsection{Spherically symmetric energy surfaces}
\label{sec:spherical_model}
\begin{figure}
\centering
\includegraphics{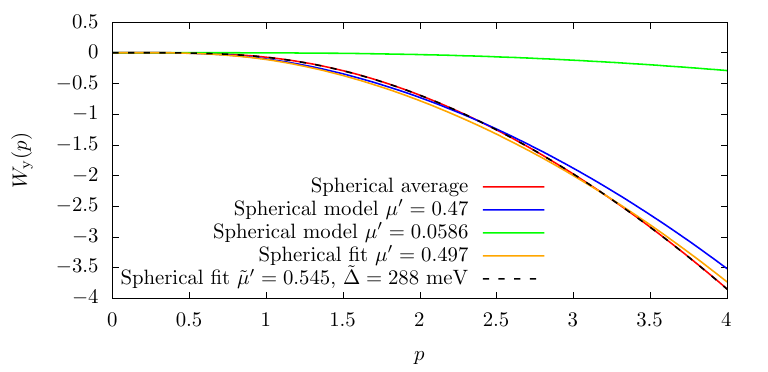}
  \caption{Comparison of the spherical average $\overline{W}_{\mathrm{y}}(p)$
    in Eq.~\eqref{eq:W_avg} (red line) with the spherical model
    $\widetilde{W}_\mathrm{y}(p)$ in Eq.~\eqref{eq:W_s} using
    $\mu'=0.0586$ (green line) and $\mu'=0.47$  (blue line). A fit of
    the spherical model to the spherical average yields $\mu'=0.497$
    (orange line). The spin-orbit splitting is fixed to the
    experimental value $\Delta=131~\mathrm{meV}$ except for the black
    dashed line where nearly perfect agreement with the spherical
    average is obtained when $\tilde\Delta=288~\mathrm{meV}$ and
    $\tilde\mu'=0.545$ are used as fit parameters for $\Delta$ and
    $\mu'$ in Eq.~\eqref{eq:W_s}.
\label{fig1:EnergySurface}}
\end{figure}

The energy surface $W_{\mathrm{y}}(\boldsymbol{p})$ for the yellow
exciton series in Eq.~\eqref{eq:Wk} has cubic symmetry $O_{\mathrm{h}}$,
i.e., the energies depend on the orientation of the momentum.
One possibility to obtain spherically symmetric energy surfaces is to
neglect the terms with cubic symmetry in the Hamiltonian~\eqref{eq:H0}
and thus to set $\nu=\delta'=\tau=\eta_1=0$ in Eq.~\eqref{eq:Wk}.
In this case the coefficients $a_3$, $b_4$, $b_5$, and $b_6$ in
Eqs.~\eqref{eq:cpq_coefs}--\eqref{eq:b_coefs} vanish.
The remaining non-zero coefficients belong to powers of
$\boldsymbol{p}^2$ and thus makes the energy surface
$W_{\mathrm{y}}(p)$ spherically symmetric.
In this special case Eq.~\eqref{eq:Wk} can be further simplified to
\begin{equation}
  \widetilde{W}_{\mathrm{y}}(p)
  = \frac{1}{4} \left(2\Delta+\frac{\mu' \gamma_1'}{m_0} p^2-
    \sqrt{4 \Delta^2+4 \Delta \frac{\mu' \gamma_1'}{m_0} p^2
      +9\frac{\mu'^2 \gamma_1'^2}{m_0^2} p^4}\right) \, .
\label{eq:W_s}
\end{equation}
Using exciton-Hartree units with $\gamma_1'/m_0=1$, 
Eq.~\eqref{eq:W_s} only depends on the spin-orbit splitting
$\Delta=131~\mathrm{meV}$ and the material parameter $\mu'=0.0586$
describing the strength of band structure contributions.
The obtained energy surface $\widetilde{W}_{\mathrm{y}}(p)$
for these parameters is shown as the green line in
Fig.~\ref{fig1:EnergySurface}.
The chosen momentum range $0\le p\le 4$ is obtained from
the semiclassical quantization condition for the angular momentum
$L=\hbar(l+\frac{1}{2})$, i.e., we can estimate that the maximum
momentum for a given $l$ in a Rydberg state is given by
\begin{equation} 
  p_{\max}=\frac{4}{2l+1} \, ,
\label{eq:MomentumRange}
\end{equation}
which yields $p_{\max}=4$ for $s$ states. 

The complete neglection of all terms with cubic symmetry may not be
justified and thus may not lead to quantitatively reasonable results.
A more reliable approach is obtained by a spherical average of the
energy surfaces in Eq.~\eqref{eq:Wk} over all orientations of the
momenta, i.e.,
\begin{equation}
  \overline{W}_{\mathrm{y}}(p) = \frac{1}{4\pi}
  \int_0^{2\pi} \int_0^{\pi} W_{\mathrm{y}}(p,\varphi_p,\vartheta_p) \mathrm{d}\Omega \, .
\label{eq:W_avg}
\end{equation}
For the computation of the integrals we use an efficient spherical
design~\cite{womersley2018efficient}.
Note that due to the $O_{\mathrm{h}}$ symmetry of the energy surfaces
the full $\Omega$ space consists of 48 fundamental regions, which are
related by symmetry properties, and thus in principle the average over
a reduced fundamental area $\Omega_{\mathrm{f}}=4\pi/48$ would be sufficient.
The energy surface $\overline{W}_{\mathrm{y}}(p)$ obtained from
Eq.~\eqref{eq:W_avg} is shown by the red line in Fig.~\ref{fig1:EnergySurface}.
Evidently, the green and red energy surfaces do not coincide, which
illustrates that ignoring of the cubic terms is indeed not justified.
However, when using $\Delta$ and $\mu'$ in Eq.~\eqref{eq:W_s} as free
parameters, the analytical form of the spherically symmetric energy
surface $\widetilde{W}_{\mathrm{y}}(p)$ can be adjusted to the
numerical surface $\overline{W}_{\mathrm{y}}(p)$ in Eq.~\eqref{eq:W_avg}. 
With the obtained fit parameters $\tilde\Delta=288~\mathrm{meV}$ and $\tilde\mu'=0.545$
the corresponding dashed black line in
Fig.~\ref{fig1:EnergySurface} nearly perfectly coincides with the red
line for the numerical surface $\overline{W}_{\mathrm{y}}(p)$.
When taking for $\Delta$ the true value of the spin-orbit coupling and
only using $\mu'$ as a fit parameter, we obtain $\mu'=0.497$, see the
orange line in Fig.~\ref{fig1:EnergySurface}.
The agreement with the red line is less perfect but still reasonable
when only using a single fit parameter.
It is interesting to note that $\mu'=0.497$ is close to the value of
$\mu'=0.47$ (see the blue line in Fig.~\ref{fig1:EnergySurface}),
obtained in Ref.~\cite{Uihlein1981Investigation} when fitting a
spherically symmetric Hamiltonian to experimental spectra.
This is a remarkable result, the small deviations of the two fit
parameters may be due to the usage of slightly different material
parameters.
The analytical functions $\widetilde{W}_{\mathrm{y}}(p)$ in
Eq.~\eqref{eq:W_s}, adjusted to the spherically symmetric energy
surfaces $\overline{W}_{\mathrm{y}}(p)$ given in Eq.~\eqref{eq:W_avg},
are the basis for the computation of quantum defects via semiclassical
EBK quantization in the following section~\ref{sec:EBK}.

\subsection{Semiclassical EBK quantization}
\label{sec:EBK}
In the spherically symmetric case the action angle variables are
characterized by spherical coordinates and a semiclassical
quantization can be obtained using the EBK quantization
conditions~\cite{einstein1917quantensatz,Brillouin1926remarques,keller1958corrected}
\begin{subequations}
\label{eq:QC}
\begin{align}
\label{eq:J_phi}
 J_\varphi &=\frac{1}{2\pi} \oint L_z \mathrm{d}\varphi=m\, , \\
\label{eq:J_theta}
 J_\vartheta &=\frac{1}{2\pi} \oint
               \sqrt{\boldsymbol{L}^2-\frac{L_z^2}{\sin^2\vartheta}}
               \mathrm{d}\vartheta=l-\left|m\right|+\frac{1}{2}\, , \\
\label{eq:J_r}
  J_r &=\frac{1}{2\pi} \oint p_r \mathrm{d}r=n_r+\frac{1}{2} \, ,
\end{align}
\end{subequations}
where the first two equations describe the quantization of the angular
momentum vector $\boldsymbol{L}$ with angular and magnetic quantum
numbers $l$ and $m$.
The radial momentum $p_r(E,r;l)$ in the third equation depends on the energy
surface $\widetilde{W}_{\mathrm{y}}(p)$, the angular momentum
$\boldsymbol{L}$, and the energy $E$, and is obtained by a numerical
root search of the equation
\begin{equation}
  E = E_{\mathrm{g}} + \frac{1}{2\mu}\left(p_r^2+\hbar^2(l+1/2)^2/r^2\right)
  +\widetilde{W}_{\mathrm{y}}\left(p=\sqrt{p_r^2+\hbar^2(l+1/2)^2/r^2}\right)-\frac{1}{r}
  \, .
\label{eq:p_r_implicit}
\end{equation}
The l.h.s.\ of the quantization condition~\eqref{eq:J_r} is evaluated
by numerical integration of $p_r=p_r(E,r;l)$ between the classical
turning points $r_{\min}$ and $r_{\max}$.
\begin{figure}
\centering
\includegraphics{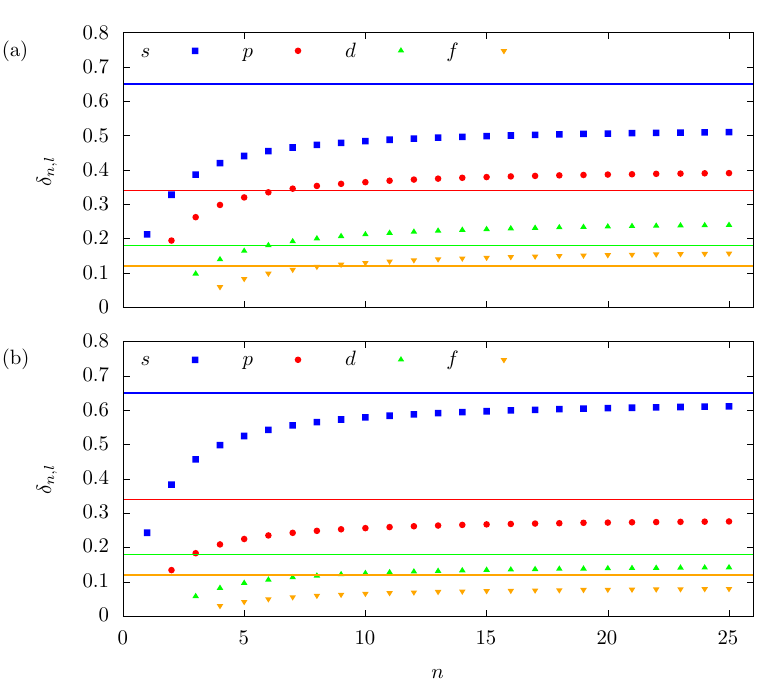}
  \caption{(a) Quantum defects $\delta_{n,l}$ for $s$, $p$, $d$, and $f$
    states obtained with the spherically symmetric energy surfaces
    $\widetilde{W}_{\mathrm{y}}(p)$ with parameters $\Delta=131~\mathrm{meV}$
    and $\mu'=0.47$.
    (b) Improved quantum defects $\delta_{n,l}$ obtained from a
    spherical model using values $\mu'_l$ optimized for different $l$
    manifolds. Here, $\mu'_l$ acts as an order parameter characterizing
    the strength of the band structure in the various $l$ manifolds.
    The solid lines mark the quantum defects in the limit of large
    principal quantum numbers $n$ given in Ref.~\cite{heckotter2017scaling}.}
\label{fig2:QuantumDefects}
\end{figure}
The quantized energies $E_{n,l}$ are finally determined, for a given
radial quantum number $n_r=n-l-1$,
from Eq.~\eqref{eq:J_r} via numerical root search.
The energy eigenvalues $E_{n,l}$ can be expressed by quantum
defects $\delta_{n,l}$ via
\begin{equation}
  E_{n,l}=E_\mathrm{g}-\frac{E_{\mathrm{Ryd}}}{(n-\delta_{n,l})^2} \, ,
\end{equation}
with the excitonic Rydberg energy $E_{\mathrm{Ryd}}$.
The spherical symmetric semiclassical model, of course, does not
provide the $l$ splitting of the $E_{n,l}$ levels in the crystal,
however, they can be compared with the mean energies of the exact
quantum mechanical $l$ manifolds, when weighting each level by its
multiplicity.

For our semiclassical calculations we use the spherically symmetric
energy surface~\eqref{eq:W_s} with $\Delta=131~\mathrm{meV}$ and
$\mu'=0.47$ to allow for direct comparisons with the results in
Ref.~\cite{Uihlein1981Investigation}.
The quantum defects $\delta_{n,l}$ up to $n=25$ for $s$, $p$, $d$, and
$f$ states are shown in Fig.~\ref{fig2:QuantumDefects}(a).
It can be seen that the quantum defects are positive and increase and 
saturate with increasing $n$, while for increasing $l$ the quantum defects 
decrease as expected.

The quantum defects obtained within the semiclassical picture are
accurate enough to reproduce the measured energies in
Ref.~\cite{Uihlein1981Investigation} up to meV precision.
In Refs.~\cite{Schoene2016Coupled,Schoene2016Deviations} experimental
spectra have been compared with an improved spherical model obtained
by using a weighted average of the band structure with cubic symmetry 
along the $[100]$ and $[111]$ directions, where analytical expressions
are known.
A comparison of these results with the quantum defects in
Fig.~\ref{fig2:QuantumDefects}(a) unveils that the semiclassical
quantum defects are overestimated in the simple spherical model for
higher angular momentum quantum numbers.
A possible reason is that the fit to the spherical average was 
performed up to the maximum allowed value of the momentum in a 
hydrogen-like $s$ state.
As can be seen from Eq.~\eqref{eq:MomentumRange} the classically
allowed momentum range decreases for increasing $l$.
The determination of the parameter $\mu'$ can be improved by considering
the probability distribution of the exciton in momentum space.
Since the energy surfaces only play a role in the quantization for the
radial direction, we assume a WKB wave function
$\psi_r \sim |p_r|^{-1/2}\, \exp(\mathrm{i}S_r/\hbar)$
in this direction, which leads to the
approximate probability distribution
\begin{equation}
  |\psi_r|^2 \sim |p_r|^{-1}=\left[p^2-\left(l+\frac{1}{2}\right)^2p^4/4\right]^{-1/2} \, .
\end{equation}
When the fitting procedure is adapted to the corresponding momentum
range and probability distribution, this in fact leads to improved
values $\mu'_l$ for different angular quantum numbers $l$.
We obtain $\mu'_s=0.51$, $\mu'_p=0.40$, $\mu'_d=0.37$, and $\mu'_f=0.34$.
Each value can be understood as an order parameter characterizing the
exciton dynamics on a specific $l$ manifold.
For increasing $l$ the spherical corrections become less important
leading to a more dominant hydrogen-like behavior.
Repeating the quantization procedure for the different models provides 
the quantum defects shown in Fig.~\ref{fig2:QuantumDefects}(b).
The $\delta_{n,l}$ are now slightly increased for $l=0$ but shifted
downwards for $l\ge 1$.
This provides a good agreement of semiclassical and quantum mechanical
values, where the quantum defects for large $n$ saturate to
$\delta_s=0.65$, $\delta_p=0.34$, $\delta_d=0.18$, and
$\delta_f=0.12$~\cite{heckotter2017scaling}, marked by the solid lines
in Fig.~\ref{fig2:QuantumDefects}(b).
In the semiclassical framework we find  
$\delta_s=0.63$, $\delta_p=0.29$, $\delta_d=0.15$, and $\delta_f=0.09$,
providing a good agreement with the literature values.
In fact  the semiclassical results are closer to the quantum mechanical 
results and experimental findings in Refs.~\cite{Schoene2016Coupled,Schoene2016Deviations}
than some values obtained by perturbative approaches~\cite{heckotter2017scaling}.

\begin{figure}
\centering
\includegraphics{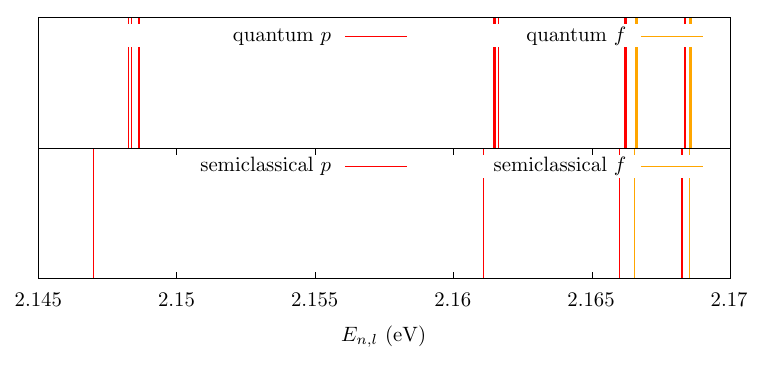}
  \caption{Comparison of energies $E_{n,l}$ of odd parity states with $n= 2$ to $5$
    obtained by numerically exact quantum calculations (top)
    with the semiclassical model using order parameters (bottom).}
\label{fig3:Comp}
\end{figure}
Finally, we compare our semiclassical results with numerically exact
eigenvalues of the Hamiltonian~\eqref{eq:H0} obtained by the method 
introduced in Ref.~\cite{Schweiner2016Impact}.
We restrict the comparison to odd-parity states with $n=2$ to $5$,
which are not affected by central-cell corrections~\cite{Schweiner2017Even}
and a coupling to the green $1s$
exciton~\cite{Rommel2021Exchange,Heckoetter2021Analysis}.
The $l$ manifolds obtained via exact diagonalization are presented in the
top of Fig.~\ref{fig3:Comp}. The splitting of different $l$ manifolds 
cannot be described within our spherical model. However, the levels
obtained by the semiclassical model (bottom) show a good agreement with 
the groups for $p$ and $f$ manifolds of the numerically exact results.
The semiclassicaly obtained energies lie systematically below the 
numerically exact eigenvalues and approach the quantum results for increasing principal quantum number $n$ 
and angular momentum quantum number $l$ as expected.
Note that our spherical model with order parameters outperforms 
the spherical models introduced previously in the literature~\cite{Uihlein1981Investigation,Schoene2016Coupled,Schoene2016Deviations},
for which the quantum defects were even bigger than our results 
and thus further away from numerically exact calculations.

\section{Conclusion and outlook}
\label{sec:conclusion}
In this article we developed and applied a spherically symmetric
semiclassical model for the description of excitons in cuprous oxide.
Therefore we used an adiabatic approach to calculate the energy
surfaces corresponding to the yellow exciton series by diagonalization
of the band-structure Hamiltonian in a basis for the quasispin and
hole spin degrees of freedom, while treating the components of the
momentum $\boldsymbol{p}$ as classical parameters.
Exploiting Kramers' theorem we could derive analytical formulas
for the energy surfaces with a cubic $O_{\mathrm{h}}$ symmetry.
The exact band structure of semiconductors such as cuprous oxide can
be computed, e.g., by application of density functional theory
(DFT)~\cite{french2008electronic}.
Since our energy surfaces are equivalent to the band structure, the
derived analytical formulas can be used for fitting the Luttinger
parameters of the Suzuki-Hensel Hamiltonian to match spin-DFT
calculations.

Spherically symmetric energy surfaces have been obtained by numerical
averaging of all orientations of the momenta, and have been fitted
with few parameters to analytical functions.
The different $l$ manifolds can be characterized by order parameters
describing the strength of the band structure contributions. 
They are the basis for an EBK quantization to obtain semiclassical
energy eigenvalues and quantum defects.
The results have been compared with earlier spherical
models~\cite{Uihlein1981Investigation,Schoene2016Coupled,Schoene2016Deviations}
and exact quantum computations for Cu$_2$O.
The same procedure can in principle also be applied to other
semiconductors with different symmetries.

The spherically symmetric model does not provide the energy splittings of
$(n,l)$ manifolds.
While the existence of a classical exciton dynamics has already been
demonstrated by using energy surfaces with the full cubic
symmetry~\cite{Ertl2022Signatures,ertl2024classical} a direct
semiclassical calculation of the energy levels is still missing.
In the future it will therefore be interesting to see whether the
semiclassical quantization scheme can be extended to correctly
reproduce the fine-structure splittings of the yellow exciton series
in cuprous oxide.
Furthermore, the analytical expressions for the energy surfaces may be
used to describe the spectra of magnetoexcitons in regions where
quantum mechanical calculations are no longer
feasible~\cite{Schweiner2017MagnetoexcitonsIn}.
To this end the closed-orbit theory developed by Du and
Delos~\cite{du1988effectI,du1988effectII} may be adopted using
energy surfaces to describe the photoabsorption spectra of
magnetoexcitons near the ionization threshold.

\bmhead{Acknowledgements}
This work was supported by Deutsche Forschungsgemeinschaft (DFG)
through Grants No.\ MA1639/16-1 and No.\ PF381/18-2 under DFG SPP1929
``Giant interactions in Rydberg systems (GiRyd).''

\bmhead{Data availability}
The data that support the findings of this study are available
from the corresponding author upon reasonable request.

\bibliography{paper}

\end{document}